\input psfig.sty
\documentstyle{cupconf}

\ifoldfss
\else
  \ifnfssone
    \newmathalphabet{\mathit}
      \addtoversion{normal}{\mathit}{cmr}{m}{it}
      \addtoversion{bold}{\mathit}{cmr}{bx}{it}
    \newmathalphabet{\mathcal}
      \addtoversion{normal}{\mathcal}{cmsy}{m}{n}
    \else
    \ifnfsstwo
    \fi
  \fi
\fi

\def\fm{\hbox{$.\!\!^{\rm m}$}}

\def\upi{\pi} 
\def\umu{\mu} 
\def\hexnumber#1{\ifcase#1 0\or1\or2\or3\or4\or5\or6\or7\or8\or9\or
 A\or B\or C\or D\or E\or F\fi }

 \font\eurmten=eurm10
 \font\eurmseven=eurm10 at 7pt
 \font\eurmfive=eurm10 at 5pt
 \newfam\eurmfam
 \textfont\eurmfam=\eurmten
 \scriptfont\eurmfam=\eurmseven
 \scriptscriptfont\eurmfam=\eurmfive
 \edef\eurm@{\hexnumber\eurmfam}
 
 \mathchardef\upi="0\eurm@19   
 \mathchardef\umu="0\eurm@16   

 \font\msxten=msam10
 \font\msxseven=msam10 at 7pt
 \font\msxfive=msam10 at 5pt
 \newfam\msxfam
 \textfont\msxfam=\msxten
 \scriptfont\msxfam=\msxseven
 \scriptscriptfont\msxfam=\msxfive
 \edef\msx@{\hexnumber\msxfam}
 \mathchardef\leqslant="3\msx@36
 \mathchardef\geqslant="3\msx@3E
 \let\le=\leqslant

\makeatletter
\ifx\CUP@mtlplain@loaded\undefined
\else
\fi
\makeatother
 \makeatletter
 \ifx\CUP@mtlplain@loaded\undefined
   \font\tenbmi=cmmib10 at 10pt
   \font\sevenbmi=cmmib10 at 7pt
   \font\fivebmi=cmmib10 at 5pt

   \newfam\bmifam
   \textfont\bmifam=\tenbmi
   \scriptfont\bmifam=\sevenbmi
   \scriptscriptfont\bmifam=\fivebmi
   
 \fi
 \makeatother
\ifnfsstwo

\fi
\ifnfssone

\fi
\ifoldfss

\fi

\mathchardef\varLambda="0103

\makeatletter
\ifx\CUP@mtlplain@loaded\undefined
\else
\fi
\makeatother
\makeatletter
\ifx\CUP@mtlplain@loaded\undefined
  \font\tenbms=cmbsy10
  \font\sevenbms=cmbsy10 at 7pt
  \font\fivebms=cmbsy10 at 5pt
  \newfam\bmsfam
  \textfont\bmsfam=\tenbms
  \scriptfont\bmsfam=\sevenbms
  \scriptscriptfont\bmsfam=\fivebms

  \edef\bsy@{\hexnumber\bmsfam}
  \mathchardef\bnabla="0\bsy@72
\fi
\makeatother
\title[Young radio-loud AGN]{Young radio-loud AGN:\\ A new sample at low redshift}

\author[I. Snellen et al.]{%
 Ignas Snellen$^1
\footnote{Present address: Institute for Astronomy, University of Edinburgh,
Royal Observatory, Edinburgh EH9 3HJ, UK}
$, Karl-Heinz Mack$^{2,3}$, Wolfgang Tschager$^{4}$ \& Richard Schilizzi$^{4,5}$}

\affiliation{$^1$Institute of Astronomy,  Madingley Road, Cambridge CB3 0HA, UK\\
$^2$ Istituto di Radioastronomia de CNR, Via Gobetti 101, I-40129 Bologna, Italy\\
$^3$ Radioastronomisches Institut der Universit\"at Bonn, Auf dem H\"ugel 71,
Bonn, Germany\\
$^4$ Leiden Observatory, P.O. Box 9513, 2300 RA, Leiden, The Netherlands\\
$^5$ Joint Institute for VLBI in Europe, Postbus 2, 7990 AA, Dwingeloo, 
The Netherlands}

\begin{document}
\ifnfssone
\else
  \ifnfsstwo
  \else
    \ifoldfss
      \let\mathcal\cal
      \let\mathrm\rm
      \let\mathsf\sf
    \fi
  \fi
\fi

\maketitle

\begin{abstract}
We report on the investigation of a complete sample of young radio-loud AGN
at $z\le 0.25$, selected using the new comprehensive radio surveys and 
the optical APM/POSS-I catalog. This new sample
will provide a unique opportunity for statistical studies of 
the early evolution of radio-loud AGN, free of cosmological evolution 
effects. In particular, their local luminosity 
function, linear-size and dynamical age distributions, may strongly constrain
possible evolution scenarios. In addition, this new sample 
is well suited for statistical studies of other important properties, like 
their X-ray and infrared emissions, optical cluster environments, and
neutral hydrogen in absorption towards the compact radio sources.
\end{abstract}

\firstsection 

\section{Young radio-loud AGN}

Cumulative evidence suggests that Gigahertz Peaked Spectrum (GPS) sources,
Compact Symmetric Objects (CSO) and Compact Steep Spectrum (CSS) sources
are young. Multi-epoch VLBI observations indicate that CSO and GPS sources
have dynamical ages of typically $10^2-10^3$ year (Owsianik \& Conway 1998;
Owsianik, Conway \& Polatidis 1998; Tschager et al. 2000), 
while radio-spectral analyses indicate that the somewhat larger CSS sources 
are up to $10^4$ years old (Murgia et al. 1999). 
It is therefore thought that these young AGN are the progenitors of large,
extended radio sources.
If so, the relatively large number of young objects suggests they
substantially decrease in radio-luminosity with time (Fanti et al. 1995; 
Readhead et al. 1996; O'Dea \& Baum 1997).

Compared with old and extended objects, young radio sources are only rarely 
found at low redshifts, e.g. only 3 GPS galaxies are found at $z<0.1$ in the
combined samples of Stanghellini et al. (1998) and Snellen et al. (1998). 
Since classes of radio sources, representing similar 
objects at different ages, are expected to have similar birth functions, 
any difference in their redshift distributions should  
be the result of a difference in their luminosity distributions.
Snellen et al. (2000) suggest that the redshift bias may be caused by 
a qualitatively different luminosity 
evolution for a radio source in the early stage of its life-cycle, 
producing a flatter collective luminosity function for young objects.
This implies that simple statistical analysis of source samples 
conducted over a large range
in cosmological epoch are unreliable and should be performed over a 
much smaller redshift range.

\section{A new sample at low redshift}

A complete sample of young radio-loud AGN 
at low redshift is an ideal tool to study the evolution 
of radio sources in more detail. Their relative proximity 
means that it is much easier to obtain key observations, like 
spectroscopic redshifts. 
Furthermore, the statistical properties of other types of 
radio sources to which the young radio sources are to be compared, 
are well determined in the local universe. 
Only since the completion of the new comprehensive radio
surveys it is possible to construct a reasonably sized sample. 
Although the objects in such a sample typically have lower
luminosities than their well-known counterparts at higher redshifts, they 
inhabit an interesting luminosity range of $P_{\rm5GHz} \sim 10^{24-26}$ W/Hz,
at and below a possible turnover in their luminosity function (Snellen et al.
2000)

 \begin{figure}[!t]
\centerline{\psfig{figure=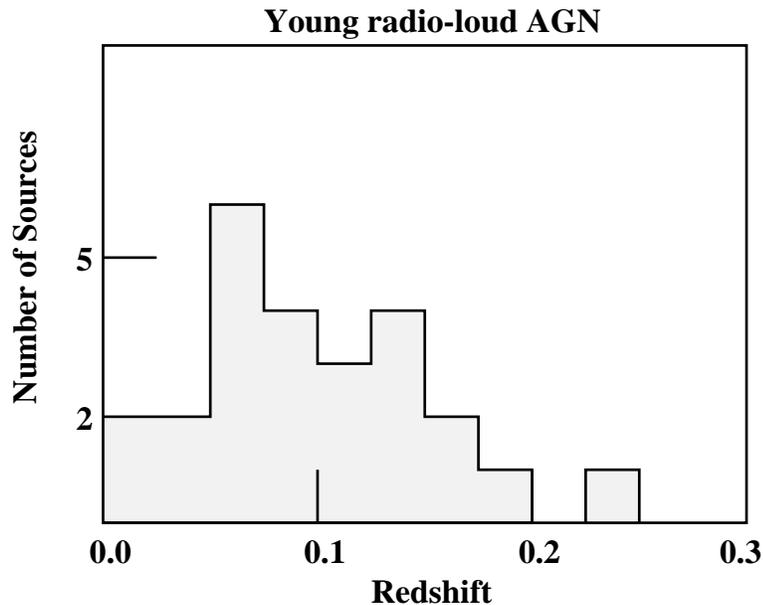,width=10cm,angle=-90} }
 \caption{The redshift distribution for the sources in our sample
($\sim 80\%$ complete).}
 \end{figure}

The combination of WENSS at 325 MHz (Rengelink et al. 1997), FIRST 
(White et al. 1997) and NVSS at 1.4 GHz (Condon et al. 1998),
GB6 at 5 GHz (Gregory et al. 1996), 
and the Automated Plate Measuring machine (APM) catalogue
of the POSS-I (McMahon \& Irwin 1991) forms the basis of our candidate 
sample of young radio sources.
Firstly, all point sources ($\theta < 2''$) were selected from the FIRST survey
with $S_{\rm 1.4GHz}>100$ mJy and correlated with the optical APM/POSS-I
catalog in the area overlapping with the WENSS survey region (1666 objects).
Only 44 of those coincide with an extended optical object in the APM/POSS-I
catalog with a red magnitude $e<16\fm5$. Although this sample is complete
for sources which are optically thin at the selection frequency, it 
may miss sources which peak at frequencies $>1.4$ GHz due to their 
compact radio morphologies. We therefore complemented the sample with 
sources from the flat spectrum CLASS survey with an 8.4 -- 5 GHz 
spectral index
such that their 1.4 GHz flux density would have been brighter than 100 mJy, 
if they were not synchrotron self-absorbed. This yielded an additional
11 sources which are optically identified with bright extended objects, and
makes the sample complete for sources with peak frequencies $<5$ GHz.

After the initial automated selection, maps from  FIRST, NVSS, WENSS 
and the Digitized Sky Survey (DSS; Lasker et al. 1990) 
centred on the positions of the 55 objects 
were checked to see whether the compact radio sources were not part of larger 
structures, and whether the optical identification was genuine.
Note that we avoided any further selection on the radio spectrum, which will 
hopefully enable us to get a better view on the questions 
whether all young radio-loud AGN are GPS sources and/or 
CSOs and/or vice versa. The remaining sample consists of 37 objects.

\section{Status and future plans}

 \begin{figure}[!t]
\centerline{\psfig{figure=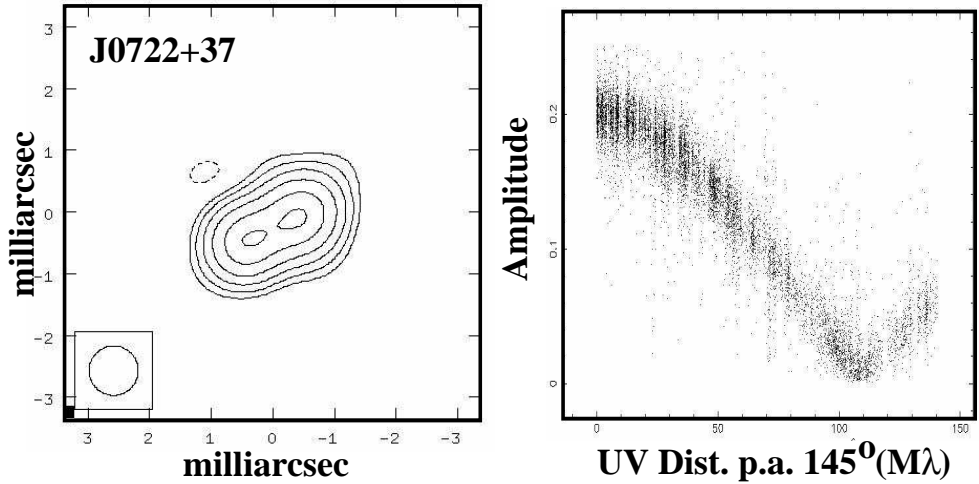,width=13cm,angle=-90}
 }
 \caption{Global VLBI data at 5 GHz for a very compact source in 
our sample. The left plot shows the amplitude of the visibility data
as function of UV distance along a position angle of 145$^\circ$. If the source
grows, the UV-minimum (currently at $\sim 110 M\lambda$) will move inwards,
which may be detectable within a few years.}
 \end{figure}

We observed the sample with the VLA at 5.0, 8.4 and 15 GHz, and 
with Effelsberg at 2.7 and 10.5 GHz (additional observations are planned 
at 32 GHz) to obtain accurate radio spectra. 
Preliminary analysis indicates that indeed all sources
are GPS or CSS sources and therefore most likely young radio sources
unaffected by Doppler boosting. Eight objects have spectroscopic redshifts
in the literature. Additional spectroscopic observations with the 
Calar Alto 2.2m telescope yielded another 17 redshifts of nearby galaxies. 
About 15 \% of the optical identifications turned out not to be 
nearby objects. The redshift-distribution of the genuine nearby
young radio sources is shown in Fig.~1 ($\sim 80 \%$ complete).

For observational strategic purposes, the sources in the sample were 
divided into 3 sub-samples according to their 
overall radio spectra: 1) The small size sub-sample, containing sources which
exhibit a turnover at frequencies $>1.4$ GHz, has been observed with the 
EVN+VLBA global VLBI array at 5 GHz. 2) The intermediate size sub-sample,
containing sources which show a spectral turnover at frequencies $<1.4$ GHz,
has been observed with the EVN only at 1.6 GHz, while observations are 
planned with MERLIN at 1.6 GHz for the `large' size sub-sample, which 
contains sources showing no sign of a spectral turnover in their overall 
radio spectrum.

Preliminary reduction of the global VLBI observations of a few
small size sources indicate that their morphologies are
consistent with them being compact doubles, as expected for these objects. 
Especially for those sources for which the flux density ratio of the two
components is near unity, multi-epoch VLBI observations will likely enable
the determination of their dynamical ages (e.g. Fig.~2). 
We believe that the dependence of age on radio luminosity and angular size, 
in combination with  their overall luminosity function and source size
distribution, will put strong constraints on possible evolution scenarios.

In addition to the interesting prospect of using this new sample 
for source evolution studies, it opens up other
possibilities as well. It provides a unique opportunity to study 
their X-ray and infrared emissions, of which very little is known at 
this moment (eg. O'Dea 1998; Fanti et al. 2000; O'Dea et al. 2000). 
In particular we look forward to studying the neutral
hydrogen gas in the centres of these galaxies 
in absorption towards the compact radio sources, which has already let to some
very interesting results in individual cases for this type of object
(eg. Conway 1995, Peck, Taylor \& Conway 1999, Pihlstr\"om et al. 1999, Conway \& Schilizzi, these proceedings).
This will now be possible for a complete sample of objects at an
unprecedented sub-parsec scale resolution.

\section*{Acknowledgements}
This research was funded by the European Commission under
contract ERBFMRX-CT96-0034 (CERES).

\end{document}